\documentclass[nofootinbib,preprintnumbers,amsmath,amssymb,aps,prc,twocolumn, showkeys,floatfix,superscriptaddress]{revtex4-1}

\usepackage{amsmath,graphics,color}
\usepackage{graphicx} 
\usepackage{booktabs, tabularx}
\usepackage{multirow,array}
\usepackage{graphics}
\usepackage{hyperref}
\usepackage{bm}
\usepackage{calligra}
\usepackage[T1]{fontenc}
\usepackage{amssymb}
\usepackage{amsfonts}
\usepackage{mathrsfs}  
\usepackage{dsfont}
\usepackage{footmisc}
\usepackage[normalem]{ulem}
\usepackage{xcolor}
\usepackage{caption}
\usepackage{subcaption}
\usepackage{ragged2e} 
\DeclareCaptionFormat{myformat}{\justifying#1#2#3}
\graphicspath{{figures/}}

\newcommand{\Tsw}{T_{\text{sw}}}

\newcommand{\taufs}{\tau_{\text{fs}}}

\DeclareMathOperator{\sign}{sign}

\begin{document}

\title{The effect of causality constraints on Bayesian analyses of heavy-ion collisions}

\author{Thiago S. Domingues}
\email{thiago.siqueira.domingues@usp.br}
\affiliation{
 Instituto de F\'isica, Universidade de S\~ao Paulo, R. do Mat\~ao, 1371, S\~ao Paulo, Brazil, 05508-090
}%

\author{Renata Krupczak}
\email{rkrupczak@physik.uni-bielefeld.de}
 \affiliation{%
 Fakultät für Physik, Universität Bielefeld, D-33615 Bielefeld, Germany
 }%

\author{Jorge Noronha}
\affiliation{%
  Illinois Center for Advanced Studies of the Universe \& Department of Physics, 
University of Illinois at Urbana-Champaign, Urbana, IL 61801-3003, USA
}%

\author{Tiago Nunes da Silva}
\email{t.j.nunes@ufsc.br}
\affiliation{%
 Departamento de F\'{i}sica, Centro de Ciências Físicas e Matemáticas, Universidade Federal de Santa Catarina, Campus Universit\'{a}rio Reitor Jo\~{a}o David Ferreira Lima, Florian\'{o}polis, Brazil, 88040-900}%

\author{Jean-François Paquet}
\email{jean-francois.paquet@vanderbilt.edu}
\affiliation{
Vanderbilt University, Nashville, TN 37235-1807
}

\author{Matthew Luzum}
\email{mluzum@usp.br}
\affiliation{
 Instituto de F\'isica, Universidade de S\~ao Paulo, R. do Mat\~ao, 1371, S\~ao Paulo, Brazil, 05508-090
}%


\date{\today}

\begin{abstract}
%
There have long been questions about the limits to the validity of relativistic fluid dynamics, and whether it is being used outside its regime of validity in modern simulations of relativistic heavy-ion collisions.  An important new tool for answering this question is a causality analysis in the nonlinear regime --- if the solutions of the evolution equations do not respect relativistic causality, they are not a faithful representation of the underlying relativistic theory (in this case, quantum chromodynamics).  Using this non-linear criterion, it has recently been shown that hydrodynamics is indeed being used outside its regime of validity in simulations, at least sometimes.  Here we explore the phenomenological implications, particularly the quantitative effects of demanding limits on acausality in modern Bayesian parameter estimation. 
We find that, while typically only a small fraction of the system's energy is initially in an acausal regime, placing strict limits on the allowed energy fraction significantly changes the preferred properties of the initial condition, which in turn alters the extracted medium properties such as bulk viscosity, where large values are no longer favored. These findings highlight the importance of developing better theoretical descriptions of the early-time, out-of-equilibrium dynamics of relativistic heavy-ion collisions.
\end{abstract}
\maketitle
\section{Introduction}
\label{sec:intro}
A key goal of the heavy-ion physics program is to quantitatively study the many-body properties of nuclear matter under extreme conditions as described by quantum chromodynamics (QCD)~\cite{CMS:2024krd, Achenbach:2023pba, Arslandok:2023utm, Shuryak:2024zrh}. The macroscopic QCD properties come from quark and gluon interactions, and we can extract information phenomenologically by colliding heavy nuclei at an ultrarelativistic regime. These relativistic heavy-ion collisions probe different regimes of nuclear matter under extremely hot and dense conditions. Numerical predictions and experimental results point to creating a fluid-like state of this matter known as the quark-gluon plasma (QGP)~\cite{Gyulassy:2004zy}. To describe the formation and space-time evolution of this fluid-like state, these collisions are typically modeled using hybrid multistage models based in large part on relativistic viscous hydrodynamics \cite{Romatschke:2017ejr, Heinz:2013th}. Since we do not have reliable first-principles calculations of many relevant properties of QCD matter, such as transport coefficients, determining these properties via comparison to experimental measurements is an essential part of the heavy-ion collision program. 

In this work, we focus on studying theoretical uncertainties and quantifying how much they can affect the extraction of the QGP medium properties, such as its transport coefficients. The hydrodynamic description of relativistic heavy-ion collisions is based on early work by Israel and Stewart \cite{Israel:1979wp}, which lead to equations of motion whose solutions have been shown to be causal \cite{Hiscock:1983zz, Olson:1990rzl, Pu:2009fj, Huang:2010sa} and stable \emph{around small perturbations in the local thermodynamic equilibrium regime}, as long as relaxation-time transport coefficients obey certain bounds. However, the regime probed by these relativistic collisions is not infinitesimally close to local equilibrium, especially at early times. Furthermore, fluid dynamics even in the ideal case is a nonlinear theory, and it is important to understand the interplay between dissipation and causality in nonlinear regime of relativistic fluids. References \cite{Bemfica:2019cop,Bemfica:2020xym, Disconzi:2023rtt} derived, for the first time, necessary and sufficient conditions that ensure causality in the nonlinear regime of Israel-Stewart-like theories undergoing general flow profiles. Nonlinearity brings about constraints on both the transport coefficients and the dissipative fluxes (such as the shear stress tensor or the bulk viscous scalar), which can affect the heavy-ion collision model-to-data comparison. The necessary conditions have been investigated in recent works \cite{ExTrEMe:2023nhy, Chiu:2021muk, Plumberg:2021bme} which showed that current numerical simulations of the QGP space-time evolution violate causality in this nonlinear regime. These studies were performed using a few sets of parameters of current \cite{Bernhard:2018hnz} heavy-ion collisions Bayesian analyses. To understand how causality bounds could affect the model-to-data comparison, it is necessary to evaluate these bounds in different regions of the parameter space to identify which combination of model parameters leads to acausality in the simulations. This causality analysis helps us understand the domain of validity of relativistic viscous hydrodynamics and, in this context, which values for the transport properties of the quark-gluon plasma are consistent with this picture. 

This paper is organized as follows. In Section \ref{sec:causal_viscous_hydro} we define our hydrodynamic model and state the conditions needed to ensure causality. Section \ref{sec.model} defines our simulation parameters, while in Section \ref{causality effects} we discuss how the causality constraints affect the extraction of the transport coefficients. Our conclusions can be found in Section \ref{sec:conclusions}. Finally, an appendix is included to provide the details of the posterior of model parameters. We use natural units $\hbar=c=k_B=1$ and a mostly minus metric signature.
\section{Relativistic Viscous Hydrodynamics and causality}
\label{sec:causal_viscous_hydro}

Relativistic hydrodynamics is a classical, macroscopic effective description of a system.  The basic degrees of freedom are those of the energy-momentum tensor $T^{\mu\nu}$ and any relevant conserved currents, which are assumed to be sufficient for determining the evolution of these same macroscopic degrees of freedom.  Here we treat the case where there are no other relevant conserved quantities besides energy and momentum.

The 10 degrees of freedom of $T^{\mu\nu}$ can be written (in the Landau frame \cite{LandauLifshitzFluids}) as
\begin{align}
    T^{\mu\nu} &= \epsilon u^\mu u^\nu - (P+\Pi)\Delta^{\mu\nu} + \pi^{\mu\nu}.
\end{align}
Here, $u^\mu$ is the fluid 4-velocity  (normalized such $u_\mu u^\mu =1$), and $\epsilon$ is the energy density in the local rest frame.
That is, they are the future-pointing timelike eigenvector and associated eigenvalue of $T^{\mu\nu}$,
\begin{align}
    u_\nu T^{\mu\nu} &= \epsilon u^\mu
\end{align}
The remaining degrees of freedom are the shear stress tensor $\pi^{\mu\nu}$, which quantifies local spatial anisotropy in the rest frame, and the bulk pressure $\Pi$ which quantifies the deviation of the rest-frame pressure from the pressure of an equilibrium system with energy $\epsilon$ defined by the equation of state, $P = P(\epsilon)$.

The evolution equations are defined by the 4 equations of conservation of energy and momentum
\begin{align}
    \partial_{\nu}T^{\mu\nu} &= 0
    \label{eq:conservation}
\end{align}
along with additional equations for the 6 viscous degrees of freedom, $\Pi, \pi^{\mu\nu}$. 

Generally, these additional hydrodynamic equations are derived as an expansion around a local equilibrium state. 
Specifically, the reference equilibrium state is the state in the rest frame defined by $u^\mu$ (i.e., the frame with zero momentum density) with energy density $\epsilon$.  If the system is in local equilibrium, then the dissipative currents vanish, $\pi^{\mu\nu} = 0 = \Pi$, and the conservation equations \eqref{eq:conservation} form a closed set of evolution equations, that of ideal hydrodynamics~\cite{Jaiswal:2016hex, Ollitrault:2012cm, Hirano:2012kj, Teaney:2009qa, Heinz:2009xj, Teaney:2009zz, Bhalerao:2008uh}.

If the system is instead only close to local equilibrium, one might write constitutive equations for the dissipative currents in terms of suitable spacetime derivatives of the original hydrodynamic variables, $\epsilon, u^\mu$. This procedure can be organized order by order in the derivative expansion, determining the dissipative fluxes. This effective field theory expansion, now called the BDNK formalism \cite{Bemfica:2017wps,Kovtun:2019hdm,Bemfica:2019knx}, is known to lead to causal, stable, and strongly hyperbolic viscous hydrodynamic theories when truncated at first order in derivatives \cite{Bemfica:2020zjp}. The BDNK approach fixes previous causality and stability issues found \cite{Hiscock:1985zz} in standard relativistic Navier-Stokes theories derived by Eckart \cite{Eckart:1940te} and Landau and Lifshitz \cite{LandauLifshitzFluids}. While this formalism has already found applications (see, e.g. \cite{Pandya:2021ief,Bemfica:2022dnk,Abboud:2023hos}), at this time there are no realistic simulations of the QGP formed in heavy-ion collisions using the BDNK equations. 


In this paper, we follow the current heavy-ion literature \cite{Romatschke:2017ejr} and use hydrodynamic equations that treat the dissipative currents as dynamical variables that obey relaxation equations, in the spirit of Israel-Stewart theory \cite{Israel:1979wp}.  These can be derived, for example, from kinetic theory (for a review, see \cite{Rocha:2023ilf}), as an expansion in quantities that represent deviations from local equilibrium --- the Knudsen number and inverse Reynolds number~\cite{ denicol2022microscopic}.  The resulting equations are~\cite{Denicol:2012cn}
%
\begin{align}
    \tau_{\Pi }\dot{\Pi}+\Pi &= -\zeta \theta -\delta _{\Pi \Pi }\Pi \theta
    + \lambda _{\Pi \pi }\pi ^{\mu \nu }\sigma_{\mu \nu }\;,
\label{relax_eqn_PI}
\\
    \tau_{\pi }\dot{\pi}^{\left\langle \mu \nu \right\rangle }+\pi ^{\mu \nu }
    &= 2\eta \sigma ^{\mu \nu }-\delta _{\pi \pi }\pi ^{\mu \nu }\theta
    +\varphi_{7}\pi _{\alpha }^{\left\langle \mu \right. }\pi ^{\left. \nu \right\rangle \alpha } \notag 
    \\
    &\ -\tau _{\pi \pi }\pi _{\alpha }^{\left\langle \mu \right. }\sigma^{\left. \nu \right\rangle \alpha }+\lambda _{\pi \Pi }\Pi \sigma ^{\mu\nu}.
\label{relax_eqn_pi}
\end{align}
Here $\dot{\Pi} = u^\lambda \partial_\lambda \Pi$, $\dot{\pi}^{\langle\mu\nu\rangle} = \Delta^{\mu\nu}_{\alpha\beta} u^\lambda \partial_\lambda \pi^{\alpha\beta}$, $\theta = \partial_\lambda u^\lambda$, and $\sigma^{\mu\nu} = \Delta^{\mu\nu}_{\alpha\beta}\partial^\alpha u^\beta$ is the shear tensor, and  $\Delta_{\alpha\beta}^{\mu\nu}$ is the transverse-traceless projector defined by
\begin{equation}
    \Delta^{\mu\nu}_{\alpha\beta} \equiv \frac{1}{2}(\Delta^{\mu}_{\alpha} \Delta^{\nu}_{\beta} + \Delta^{\nu}_{\alpha} \Delta^{\mu}_{\beta}) - \frac{1}{3}\Delta^{\mu\nu}\Delta_{\alpha\beta},
\label{eq:delta_munu_alphabeta}
\end{equation}
while the projector orthogonal to the flow is
\begin{equation*}    
    \Delta_{\mu\nu} \equiv g_{\mu\nu} - u_{\mu}u_{\nu},
\end{equation*}
where $g_{\mu\nu}$ is the spacetime metric. The shear stress tensor is symmetric, $\pi_{\mu \nu} = \pi_{\nu \mu}$, traceless, $\pi^{\mu}_{\mu} = 0$, and orthogonal to the fluid's 4-velocity, $\pi^{\mu \nu}u_{\mu} = 0$.
The transport coefficients represent properties of the underlying dynamics, and their value and dependence on local temperature are specific to each particular fluid: $\tau_\Pi$, $\zeta$, $\delta_{\Pi\Pi}$, $\lambda_{\Pi\pi}$, $\tau_\pi$, $\eta$, $\delta_{\pi\pi}$, $\phi_7$, $\tau_{\pi\pi}$, $\lambda_{\pi\Pi}$.

Hydrodynamics is typically expected to be valid when non-equilibrium corrections are small. Note, however, that since even an ideal fluid is a nonlinear system, near-equilibrium behavior does not imply that the system's dynamics is linear. Furthermore, hydrodynamic behavior may be meaningful even when dissipative fluxes are large, see \cite{Heller:2015dha,Denicol:2021wod,Soloviev:2021lhs,Gavassino:2023xkt}. This makes it difficult to determine with certainty what is the domain of validity of hydrodynamics unless one can solve the underlying microscopic theory. This is usually not the case, especially for QCD and simulations of relativistic heavy-ion collisions. Therefore, if one assumes that a fluid description is valid and that relativistic effects are relevant, causality is a \emph{necessary} condition to ensure consistency of the framework (i.e., the system of nonlinear PDEs can be hyperbolic and have a well-posed initial value problem) and stability of near-equilibrium disturbances \cite{Bemfica:2020zjp,Gavassino:2021owo}. Thus, violation of relativistic causality can be effectively used as a criterion that indicates the breakdown of a given hydrodynamic description.
\subsection{Linear causality conditions}
\label{subsec:linear_conditions}
In the simplest causality analysis, one considers a uniform system in thermal equilibrium with an infinitesimal perturbation \cite{Hiscock:1983zz,Olson:1990rzl,Romatschke:2009im}.  In this case, one can linearize the hydrodynamic equation to derive a condition on the transport coefficients of the underlying fluid for the hydrodynamic evolution to respect relativistic causality, 
\begin{equation}
    n_\text{static} \equiv c_s^2 + \frac{4}{3} \frac{\eta}{\tau_\pi (\epsilon + P)} + \frac{\zeta}{\tau_\Pi (\epsilon + P)} \leq 1.
    \label{eq:linear_condition}
\end{equation}
Here $c_s$ is the sound speed given by $\sqrt{dP/d\epsilon}$.  The relevant transport coefficients that enter in a linear analysis are the shear viscosity $\eta$ and bulk viscosity $\zeta$ and their respective relaxation times\footnote{We note that another necessary condition is $0\leq \frac{\eta}{\tau_\pi (\epsilon+P)}\leq 1$, which is less stringent than our \eqref{eq:linear_condition} given that $c_s \in [0,1]$ and $\zeta>0$ in this work.} $\tau_\pi$ and $\tau_\Pi$ \cite{Olson:1989ey}. 
Here we can see confirmation of the natural expectation that if the relaxation times are too short, the local bulk pressure or shear tensor can relax so quickly that signals propagate faster than the speed of light, thereby signaling causality violation.  
From our discussion above, any realistic relativistic viscous fluid should have transport coefficients that satisfy this condition.
\subsection{\textbf{Nonlinear causality conditions}}
\label{subsec:nonlinear_causality}
The hydrodynamic equations constitute a nonlinear set of PDEs and, as such, causality has to be established also in the full nonlinear regime. A more comprehensive, fully nonlinear causality analysis of Israel-Stewart-like theories under general flow conditions was only recently performed in \cite{Bemfica:2019cop,Bemfica:2020xym}. In \cite{Bemfica:2020xym}, both necessary and sufficient conditions were derived for the hydrodynamic equations with shear and bulk viscosity effects to respect relativistic causality. Here, following \cite{Chiu:2021muk}, the necessary conditions for causality take the form
\allowdisplaybreaks
\begin{subequations}
\begin{align}
    n_1 &\equiv \frac{2}{C_\eta}+\frac{\lambda_{\pi\Pi}}{\tau_\pi}\frac{\Pi}{\varepsilon+P}-\frac{\tau_{\pi\pi}}{2\tau_\pi}\frac{|\Lambda_1|}{\varepsilon+P} \geq 0,
    \label{eq:causal_n1} \\
    n_2 &\equiv 1 - \frac{1}{C_\eta}+\left(1-\frac{\lambda_{\pi\Pi}}{2\tau_\pi}\right)\frac{\Pi}{\varepsilon+P}-\frac{\tau_{\pi\pi}}{4\tau_\pi}\frac{\Lambda_3}{\varepsilon+P} \geq 0,
    \label{eq:causal_n2} \\
    n_3 &\equiv \frac{1}{C_\eta}+\frac{\lambda_{\pi\Pi}}{2\tau_\pi}\frac{\Pi}{\varepsilon+P}-\frac{\tau_{\pi\pi}}{4\tau_\pi}\frac{\Lambda_3}{\varepsilon+P} \geq 0,
    \label{eq:causal_n3} \\
    n_4 &\equiv 1 - \frac{1}{C_\eta} + \left(1-\frac{\lambda_{\pi\Pi}}{2\tau_\pi}\right)\frac{\Pi}{\varepsilon+P} \nonumber \\
    &\quad + \left(1-\frac{\tau_{\pi\pi}}{4\tau_\pi} \right) \frac{\Lambda_a}{\varepsilon+P}-\frac{\tau_{\pi\pi}}{4\tau_\pi}\frac{\Lambda_d}{\varepsilon+P}\geq0,\, (a\neq d)
    \label{eq:causal_n4} \\
    n_5 &\equiv c^2_s + \frac{4}{3}\frac{1}{C_\eta} + \frac{1}{C_\zeta} + \left(\frac{2}{3}\frac{\lambda_{\pi\Pi}}{\tau_\pi} + \frac{\delta_{\Pi\Pi}}{\tau_\Pi} + c^2_s \right)\frac{\Pi}{\varepsilon+P} \nonumber \\
    &\quad + \left( \frac{3\delta_{\pi\pi} + \tau_{\pi\pi}}{3\tau_\pi} + \frac{\lambda_{\Pi\pi}}{\tau_\Pi} + c^2_s\right) \frac{\Lambda_1}{\varepsilon + P} \geq 0,
    \label{eq:causal_n5} \\
    _6 &\equiv 1 - \left(c^2_s + \frac{4}{3}\frac{1}{C_\eta} + \frac{1}{C_\zeta}\right) \nonumber \\
    &\quad + \left(1 - \frac{2}{3}\frac{\lambda_{\pi\Pi}}{\tau_\pi} - \frac{\delta_{\Pi\Pi}}{\tau_\Pi} - c^2_s\right)\frac{\Pi}{\varepsilon+P} \nonumber \\
    &\quad + \left(1 - \frac{3\delta_{\pi\pi} + \tau_{\pi\pi}}{3\tau_\pi} - \frac{\lambda_{\Pi\pi}}{\tau_\Pi} - c^2_s\right)\frac{\Lambda_3}{\varepsilon + P} \geq 0.
    \label{eq:causal_n6}
\end{align}
\label{eq:causal_n}
\end{subequations}
where $C_\eta = \tau_\pi(\varepsilon + P)/\eta$ and $C_\zeta = \tau_\Pi(\varepsilon + P)/\zeta$ are the dimensionless coefficients for the ratios of the shear and bulk relaxation times that appear in the linearized condition \eqref{eq:linear_condition}. The condition $n_4$ must hold for $a,d = 1,2,3$. In this more general scenario, the causality conditions depend on all additional transport coefficients as well as the bulk viscous pressure $\Pi$ and the three nonzero eigenvalues of the shear tensor, $\Lambda_{a}$, $a =1,2,3$. The eigenvalues are ordered by value such that $\Lambda_1 \leq \Lambda_2 \leq \Lambda_3$.  Note that, since the shear tensor is defined to be traceless, $\Lambda_1$ is negative while $\Lambda_3$ is positive. The conditions above are necessary, i.e., if they are violated it means that causality is certainly violated. Ref.\ \cite{Bemfica:2020xym} also determined sufficient conditions for causality, meaning that if they are satisfied, for sure causality is satisfied (however, if they are violated, this does not mean acausality). In this work, we focus solely on the necessary conditions for causality shown explicitly above.

%


%
\section{JETSCAPE SIMULATION FRAMEWORK }
\label{sec.model}

In this work, we use as a baseline the Bayesian parameter estimations performed by the JETSCAPE Collaboration in References \cite{JETSCAPE:2020mzn, JETSCAPE:2020shq}.  By comparing numerical results to a large set of experimental data, constraints can be put on unknown parameters of the model used to describe the system.  The main elements of the simulation model used in these works are
%
%
\begin{itemize}
    \item $\mathrm{T}_{\mathrm{R}} \mathrm{ENTo}$ \cite{Moreland:2014oya} to generate the system's initial energy density;
    \item Free-streaming of the initial $\mathrm{T}_{\mathrm{R}} \mathrm{ENTo}$ profile using \cite{Liu:2015nwa};
   \item Viscous hydrodynamic evolution using the MUSIC code \cite{Schenke:2010nt};
   \item Particlization based on the Cooper-Frye formula using the frzout code  \cite{PhysRevD.10.186};
   \item Boltzmann evolution and decays of hadrons using the UrQMD code \cite{Bleicher:1999xi}.
\end{itemize}

Here we briefly summarize these stages, referring the reader to Refs.~\cite{JETSCAPE:2020mzn, JETSCAPE:2020shq} for more details.  A full list of model parameters and their prior range are shown in Table \ref{tb:prior_table}.

\begin{table*}
{%
\footnotesize
\begin{tabular}{||p{0.2\linewidth}|p{0.11\linewidth}|p{0.15\linewidth}||p{0.2\linewidth}|p{0.08\linewidth}|p{0.15\linewidth}||}
Norm. Pb-Pb 2.76 TeV & $N$[2.76 TeV] & {[}10, 20{]} & temperature of $(\eta/s)$ kink & $T_{\eta}$ & {[}0.13, 0.3{]} GeV \\
Norm. Au-Au 200 GeV & $N$[0.2 TeV] & {[}3, 10{]} & $(\eta/s)$ at kink & $(\eta/s)_{\rm kink}$ & {[}0.01, 0.2{]} \\
generalized mean & $p$ & {[}--0.7, 0.7{]} & low temp. slope of $(\eta/s)$ & $a_{\text{low}}$ & {[}--2, 1{]} GeV$^{-1}$ \\
nucleon width & $w$ & {[}0.5, 1.5{]} fm & high temp. slope of $(\eta/s)$ & $a_{\text{high}}$ & {[}--1, 2{]} GeV$^{-1}$ \\
min. dist. btw. nucleons & $d_{\text{min}}^3$ & {[}0, 1.7$^3${]} fm$^3$ & shear relaxation time factor & $b_{\pi}$ & {[}2, 8{]} \\
multiplicity fluctuation & $\sigma_k$ & {[}0.3, 2.0{]} & maximum of $(\zeta/s)$ & $(\zeta/s)_{\text{max}}$ & {[}0.01, 0.25{]} \\
free-streaming time scale & $\tau_R$ & {[}0.3, 2.0{]} fm/$c$ & temperature of $(\zeta/s)$ peak & $T_{\zeta}$ & {[}0.12, 0.3{]} GeV \\
free-streaming energy dep. & $\alpha$ & {[}--0.3, 0.3{]} &  width of $(\zeta/s)$ peak & $w_{\zeta}$ & {[}0.025, 0.15{]} GeV \\
particlization temperature & $\Tsw$ & {[}0.135, 0.165{]} GeV & asymmetry of $(\zeta/s)$ peak & $\lambda_{\zeta}$ & {[}--0.8, 0.8{]}
\end{tabular} 
}
\caption{
Model parameters and prior ranges from Ref.~\cite{JETSCAPE:2020mzn}.  Before imposing causality constraints the prior is taken to be uniform within this range of values.
}
\label{tb:prior_table}
\end{table*}
\subsection{Pre-hydrodynamic stage}
\label{subsec:pre_hydro}
\subsubsection{$\mathrm{T}_{\mathrm{R}} \mathrm{ENTo}$ Ansatz}
\label{sub_subsec:trento}
$\mathrm{T}_{\mathrm{R}} \mathrm{ENTo}$ is a model that parametrizes the transverse distribution of deposited energy in the earliest stage of a collision.  It is designed to be flexible, such that it can approximate the energy deposition of different physical models with different parameter values.

The first step is to sample the positions of nucleons within each colliding nucleus according to a Woods-Saxon distribution.  The sampling is done so that the (3D) radii and polar angles are sampled independently, while the azimuthal angle is sampled to ensure that no two nucleons have a distance of less than $d_{\rm min}$ from each other, to roughly approximate the effects of short-range nuclear interactions.  This exclusion volume $d_{\rm min}^3$ is taken as a free parameter to be determined from measurements.

Each nucleon is then associated with a Gaussian profile of width $w$, which is a free parameter that has two main functions.   First, it determines the impact-parameter-dependent criterion to determine whether each potential target-projectile nucleon pair suffers an inelastic collision, and second, it determines the transverse extent of deposited energy.
%

The nucleons that suffer no collisions are excluded, and the remaining ``participants'' are used to construct the thickness function for each nucleus as a sum of the Gaussian densities of each nucleon.  Each Gaussian is multiplied by a factor that is drawn randomly from a Gamma distribution of standard deviation $\sigma_k$, another free parameter in the analysis.  This is meant to represent fluctuations in the particle production processes, for example as observed in proton-proton collisions.

The initial energy density, then, is proportional to a generalized mean of the thickness functions of the two colliding nuclei

\begin{align}
    \tau \epsilon(\mathbf{x}_\perp) &= N \left(\frac{T^p_A(\mathbf{x}_\perp) + T^p_B(\mathbf{x}_\perp)}{2}\right)^{1/p}\\
    T_{A/B}(\mathbf{x}_\perp) &= \sum_{i\in A/B} \gamma_i \rho(\mathbf{x}_\perp{-}\mathbf{x}_{i,\perp})\\
    \rho(\mathbf{x}_\perp) &= \int_{-\infty}^{\infty} \frac{dz}{\left(2\pi w^2\right)^{3/2}}\exp\left(-\frac{\mathbf{x}_\perp^2+z^2}{2w^2}\right),
\end{align}
where the additional parameters are $p$, which determines the type of average taken, and an overall normalization factor $N$.

\subsubsection{Free-streaming}
\label{sub_subsec:free-streaming}
Pre-hydrodynamic evolution is modeled as a sort of free-streaming expansion, with massless particles propagating in a purely transverse direction with an initially isotropic distribution.  The energy-momentum tensor at time $\tau_{\rm fs}$ is thus
\begin{align}
\label{eq:Tmn_fs}
    T^{\mu\nu}(t,\boldsymbol{x})   
    &= \int \frac{d\Omega_p}{4\pi}\, \hat{p}^{\mu} \hat{p}^{\nu}\, \epsilon\bigl( \boldsymbol{x}{-}\boldsymbol{v}\left(t{-}t_{\rm fs}\bigr)\right),
\end{align}
where $\Omega_p$ is the solid angle in momentum space and $\hat p^\mu = p^\mu/E$.

The free-streaming time in a given event depends on two free parameters
\begin{equation}
   \taufs = \tau_R \left(\frac{\langle \bar\epsilon \rangle}{\bar\epsilon_R}\right)^{\alpha}.
   \label{eq:taufs}
\end{equation}
where 
$\langle \bar\epsilon \rangle$ is the average initial energy density in the transverse plane
\begin{equation}
    \langle \bar\epsilon \rangle \equiv \frac{\int d^2x_\perp\, {\bar\epsilon}^2(\mathbf{x}_\perp)}{\int d^2x_\perp\, \bar\epsilon(\mathbf{x}_\perp)}.
\label{eq:eps_ave}    
\end{equation}
and we choose $\bar\epsilon_R = 4.0$\, GeV/fm$^2$ as a reference scale, noting that the resulting posteriors for $\alpha$ and $\tau_R$ depend on this choice.

This energy-momentum tensor is used as the initial condition for hydrodynamic evolution.  It is important to note that the massless degrees of freedom ensure a vanishing trace
\begin{align}
    T^\mu_{\phantom{\mu}\mu} = 0,
\end{align}
while the equilibrium equation of state used in the hydrodynamic phase is non-conformal, $P(\epsilon) < \epsilon/3$.   The resulting difference in pressure is absorbed in the bulk pressure, which as a result is determined only by the local energy density and not by any other aspect of the local conditions or  evolution history
\begin{align}
    \Pi = \frac \epsilon 3 -  P(\epsilon).
\end{align}
This choice implies that the bulk pressure is always positive $\Pi>0$, and potentially large, at the onset of hydrodynamic evolution.

The results of this work depend crucially on the model used, and a more realistic, non-conformal pre-hydrodynamic model may well give different results \cite{daSilva:2022xwu}.
\subsection{Hydrodynamic stage}
\label{subsec:hydro_stage}
We evolve the energy-momentum tensor of the system using second-order relativistic viscous hydrodynamics (see the detailed discussion in Section \ref{sec:causal_viscous_hydro}) whose numerical implementation is given by the dissipative fluid dynamics code MUSIC \cite{hydro_code}. 
Since we are interested in describing systems in the midrapidity region we can assume (2+1)-dimensional dynamics with longitudinal boost-invariance. In this stage, we employ a lattice QCD equation of state with a temperature-dependent parametrization for the first-order transport coefficients. In a system without conserved charges, the transport coefficients are only temperature-dependent,
and both the specific shear and bulk viscosities are parametrized as the ratios of shear and bulk viscosity to
entropy density — the unitless specific viscosities. 

The specific shear viscosity is parametrized by a piecewise linear function:     
\begin{equation}
\label{eq:positivity}
    \frac{\eta}{s}(T) = \max\left[\left.\frac{\eta}{s}\right\vert_{\rm lin}\!\!\!(T),0\right],
\end{equation}
with
\begin{eqnarray}
    \left.\frac{\eta}{s}\right\vert_{\rm lin}\!\!\!(T) &=& a_{\rm low}\, (T{-}T_{\eta})\, \Theta(T_{\eta}{-}T)+ (\eta/s)_{\rm kink}
\nonumber\\
    && +\, a_{\rm high}\, (T{-}T_{\eta})\, \Theta(T{-}T_{\eta}).
\end{eqnarray}
Where here, $T_{\eta}$ is the position of the inflection point at or above the deconfinement transition \cite{Niemi:2015qia}, $(\eta/s)_{\rm kink}$ is the value of $\eta/s$ at this point, $a_{\rm low}$ and $a_{\rm high}$ are slopes below and above the inflection point, respectively. Both slopes can assume negative and positive values.

The specific bulk viscosity is described by a skewed Cauchy distribution:
\begin{eqnarray}
    \frac{\zeta}{s}(T) &=& \frac{(\zeta/s )_{\max}\Lambda^2}{\Lambda^2+ \left( T-T_\zeta\right)^2},  \label{eq:bulk_parametrization} \\
    \Lambda &=& w_{\zeta} \left[1 + \lambda_{\zeta} \sign \left(T{-}T_\zeta\right) \right]\nonumber.
\end{eqnarray}
We are assuming that a single peak is near the deconfinement temperature \cite{Kharzeev:2007wb, Karsch:2007jc, NoronhaHostler:2008ju, Rose:2020lfc, Arnold:2006fz}.
Here, $T_\zeta$ is the peak position, $(\zeta/s)_{\max}$ is the value of $\zeta/s$ at this point, $w_\zeta$ is the width of the Cauchy distribution, and $\lambda_{\zeta}$ is its skewness.  
Both shear and bulk viscosities must be positive-definite due to the second law of thermodynamics. 
The viscous currents (i.e. the shear stress $\pi^{\mu \nu}$ and bulk viscous pressure $\Pi$) have their relaxation equations where terms of first and second-order appear to describe the near-equilibrium dynamics. The second-order transport coefficients entering into the second-order hydrodynamic theory are $\delta _{\Pi \Pi }$, $\lambda _{\Pi \pi }$, $\delta _{\pi \pi }$, $\varphi _{7}$, $\tau _{\pi \pi }$, and $\lambda_{\pi\Pi }$. The other two second-order transport coefficients are the shear and bulk relaxation times $\tau _{\pi }$ and $\tau _{\Pi }$ which play a central role in Israel-Stewart-like theories. 
The JETSCAPE framework parametrizes the shear and bulk viscosities, $\eta$ and $\zeta$, as functions of temperature. However, we have less theoretical guidance for the second-order transport coefficients, and following\cite{JETSCAPE:2020mzn} all the second-order transport coefficients are related to first-order ones by using parameter-free relations derived from kinetic theory \footnote{See \cite{Rocha:2024rce} for calculations using different assumptions for the collision kernel of kinetic theory.} \cite{Denicol:2012cn},
\begin{subequations}
\begin{align}
    \frac{\delta_{\Pi \Pi}}{\tau_{\Pi}} &= \frac{2}{3} \\
    \frac{\delta_{\pi \pi}}{\tau_{\pi}} &= \frac{4}{3} \\
    \frac{\tau_{\pi \pi}}{\tau_{\pi}} &= \frac{10}{7} \\
    \frac{\lambda_{\pi \Pi}}{\tau_{\pi}} &= \frac{6}{5} \\
    \frac{\lambda_{\Pi \pi}}{\tau_{\Pi}} &= \frac{8}{5} \left( \frac{1}{3}-c_s^2 \right),
\end{align}
\label{eq:transport_coeff}
\end{subequations}
%

The shear relaxation time $\tau_{\pi}$ is parametrized
by following temperature dependence: 
\begin{equation}
    T \tau_\pi(T)= b_{\pi}\frac{\eta}{s}(T)
    \label{eq:tau_pi}
\end{equation}
where $b_\pi$ is a model parameter, a constant that we can constrain. The previous work \cite{JETSCAPE:2020mzn}, used the linearized causality bound \cite{Pu:2009fj} requiring $b_\pi{\,\ge\,}(4/3)/(1{-}c_s^2){\,\ge\,}2$ motivating the prior range explored for $b_\pi$ from a variety of weakly and strongly coupled theories varying $b_\pi$ between ${\sim}2$ and ${\sim}6$. The specific bulk viscosity is given by
\begin{equation}
    \tau_{\Pi} = b_{\Pi} \frac{\zeta}{\left(\frac{1}{3} - c_s^2\right)^2 (\epsilon + p)}
    \label{eq:tau_Pi}
\end{equation}
where $b_{\Pi} = 1/14.55$ \cite{denicol2014transport}.
The bulk relaxation time \cite{Denicol_2009} $\tau_{\Pi}$ represents a time scale to the bulk pressure $\Pi$ relax to its Navier-Stokes value $\Pi_{NS} = - \zeta \theta$, where $\theta$ is the expansion rate.
Another important ingredient that enters the hydrodynamic stage is the equation of state (EoS). Here, we used an equation of state provided by matching the HotQCD EoS \cite{bazavov2019chiral} with a hadronic resonance gas~\cite{Bernhard:2018hnz}\footnote{\url{https://github.com/j-f-paquet/eos_maker}}.
\subsection{Post-hydrodynamic stage}
\label{subsec:post-hydro}
%
%
Eventually, one must switch from a fluid description to a description in terms of the particles that are finally measured.   This switching of descriptions is done on a spacetime hypersurface of constant temperature $T_{\rm sw}$.  

The distribution of particles exiting this switching surface is then evolved according to the Boltzmann equation by the model \cite{SMASH:2016zqf}.   Final observables are constructed according to the respective experimental procedure from the hadrons that remain after the last scattering or decay has occurred.

\section{Imposing limits on causality}\label{causality effects}
When designing a Bayesian parameter estimation, an essential aspect is to choose a prior probability distribution, which is a statement about our knowledge of model parameters \textit{before} comparing to the experimental data.  In the absence of any particular theoretical knowledge of parameter values, this is often taken to be a uniform distribution for each parameter (within some finite range), even though this choice remains an arbitrary one.

However, knowing which parameter values violate (or are likely to violate) relativistic causality during system evolution, we can naturally use this information to inform the Bayesian inference via the prior distribution.

\subsection{Linear causality conditions}

With the parameterization of the shear and bulk relaxation times given by Equations \eqref{eq:tau_pi} and \eqref{eq:tau_Pi}, the linearized causality condition \eqref{eq:linear_condition} simplifies to
\begin{equation}
    n_\text{static} \equiv c_s^2 + \frac{4}{3b_{\pi}} + \frac{(1/3 -c_s^2)^2}{b_{\Pi}} \leq 1 ~,
    \label{linear condition parameters}
\end{equation}
which involves only the relaxation time prefactor constants $b_\Pi, b_\pi$ and the (temperature-dependent) speed of sound $c_s$.  

Since hydrodynamics should certainly be a valid description when arbitrarily close to thermal equilibrium, the relaxation times of any realistic theory should satisfy this condition (or more precisely Eq.~\eqref{eq:linear_condition}).

\begin{figure}
    \centering
    \includegraphics[width=\linewidth]{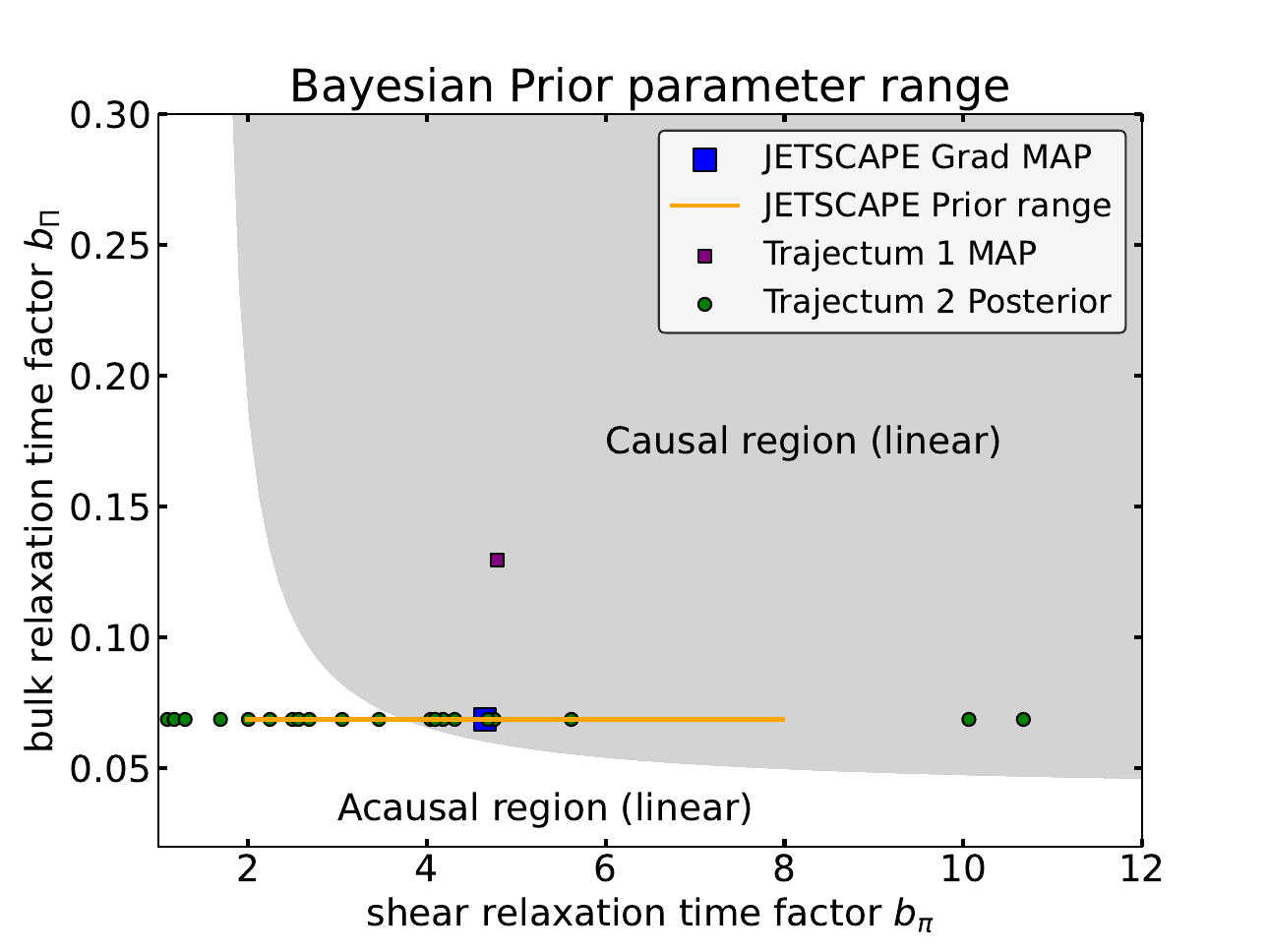}
    \caption{Prior range for relaxation times \eqref{eq:linear_condition} in recent Bayesian analyses. The entire plot corresponds to the prior of Ref.~\cite{Nijs:2020roc} (labeled as Trajectum 1), where both parameters were varied. In contrast, Ref.~\cite{Nijs:2021clz} (labeled as Trajectum 2) and JETSCAPE \cite{JETSCAPE:2020mzn} kept the bulk relaxation factor fixed to $b_\Pi = 1/14.55$ and varied only $b_\pi$. The shaded region is allowed by the linear causality condition \eqref{eq:linear_condition}. Also shown are 20 random samples of the Trajectum 2 posterior \cite{Nijs:2021clz} and the Maximum A Posteriori (MAP) point for JETSCAPE Grad \cite{JETSCAPE:2020mzn} and Trajectum 1 \cite{Nijs:2020roc}. 
    }
    \label{fig:Bayesian_prior_range}
\end{figure}

Several recent Bayesian analyses have included one or both of these prefactors in their parameter estimation \cite{Nijs:2020roc, Nijs:2021clz, JETSCAPE:2020mzn}.  In Fig.~\ref{fig:Bayesian_prior_range}, we show the corresponding prior ranges for $b_\Pi, b_\pi$.  The entire plot corresponds to the prior of Ref.~\cite{Nijs:2020roc} (labeled as Trajectum 1), where both parameters were varied.  In contrast, Ref.~\cite{Nijs:2021clz} (labeled as Trajectum 2) kept the bulk relaxation factor fixed to $b_\Pi = 1/14.55$ and varied only $b_\pi$, as did the JETSCAPE analyses \cite{JETSCAPE:2020mzn}.

Interestingly, we can see that a significant portion of the prior is in the unphysical regime that violates the linear causality condition (shown as the white region) in all of these analyses.   Also shown in Fig.~\ref{fig:Bayesian_prior_range} are the Maximum A Posteriori (MAP) point for Trajectum 1 and JETSCAPE, which is the parameter value that maximizes the likelihood after comparison to experimental constraints.   These lie in the causal region, although this is perhaps a coincidence since the posterior for these parameters is largely uniform, shown here with 20 samples of the Trajectum 2 posterior.

Thus, an imposition of linear causality on the models has a significant effect on not only the prior but also the posterior. In the JETSCAPE analysis, 31\% of the posterior is in a regime that violates the linearized causality condition.  However, we find that disallowing this violation by restricting the relaxation time does not significantly affect other extracted parameters, due to the low sensitivity of the included observables to the shear relaxation time~\cite{Everett:2021ruv}.  In effect, experimental data do not demand that the shear relaxation time be sufficiently large so as to satisfy the long-known linear causality conditions, but neither does the inclusion of unphysical values in the prior bias the extraction of other model parameters of interest in a non-negligible way.

In realistic simulations, the system does not necessarily remain perturbatively close to thermal equilibrium, and so it is of great interest to know the effects of violating the full (nonlinear) causality conditions, which we discuss in the following.
\subsection{Full nonlinear causality conditions}
\subsubsection{Priors}
As with the linear conditions in the previous section, even before performing simulations one can analyze the model parameter space to identify when causality violation is likely to happen, and use that information to choose priors for the Bayesian calibration.  However, in this nonlinear case, it is more complicated, since in addition to the local speed of sound, which has a range of possible values,
the conditions now depend on local deviations from equilibrium as encoded in bulk pressure $\Pi$ and eigenvalues of the shear stress tensor $\pi$.

Using the standard parameterizations of transport coefficients~\eqref{eq:transport_coeff}, the six necessary causality conditions~\eqref{eq:causal_n} simplify to the following quantities, which must be non-negative for the hydrodynamic equations to respect causality
\begin{align}
    n_1 &= \frac{2}{C_{\eta}}+\frac{6}{5}\frac{\Pi}{\varepsilon+P}-\frac{5}{7}\frac{|\Lambda_1|}   {\varepsilon+P} \geq 0\\
    n_2 &= 1 - \frac{1}{C_{\eta}}+\frac{2}{5}\frac{\Pi}{\varepsilon+P}-\frac{5}{14}\frac{\Lambda_3}{\varepsilon+P} \geq 0\\
    n_3 &= \frac{1}{C_{\eta}}+\frac{3}{5}\frac{\Pi}{\varepsilon+P}-\frac{5}{14}\frac{\Lambda_3}{\varepsilon+P} \geq 0\\
    n_4 &= 1 - \frac{1}{C_{\eta}}+ \frac{2}{5} \frac{\Pi}{\varepsilon+P} + \frac{9}{14} \frac{\Lambda_a}{\varepsilon+P}-\frac{5}{14}\frac{\Lambda_d}{\varepsilon+P}\geq 0\\
    n_5 &= c_s^2 + \frac{4}{3C_{\eta}}+ \frac{1}{C_{\zeta}} + \left(\frac{22}{15} + c_s^2\right) \frac{\Pi}{\varepsilon+P}\nonumber \\
    &\qquad + \left(\frac{38}{21} + \frac{8(1/3 - c_s^2)}{5} + c_s^2\right) \frac{\Lambda_1}{\varepsilon+P} \geq0 \label{eq:n5} \\
    n_6 &= 1 - c_s^2 - \frac{4}{3C_{\eta}} - \frac{1}{C_{\zeta}} + \left(-\frac{7}{15} - c_s^2\right) \frac{\Pi}{\varepsilon+P} \nonumber \\
    &\qquad + \left(-\frac{17}{21} - \frac{8(1/3 - c_s^2)}{5} - c_s^2\right) \frac{\Lambda_3}{\varepsilon+P} \geq0 \label{eq:n6}
\end{align}
These quantities can depend crucially on the pre-hydrodynamic evolution of the system and can have a large variation in spacetime and event-to-event.
Nevertheless, we can estimate their likely range and identify the most stringent conditions by using the prior range of $b_\pi \equiv C_\eta$ and estimating the reasonable range of the inverse Reynolds numbers following~\cite{Chiu:2021muk}: $0 \leq R_{\Pi} \equiv \frac{\Pi}{\varepsilon+P} \leq 1$ and $-1 \leq \frac{\Lambda_1}{\varepsilon+P} \leq 0$, $0 \leq \frac{\Lambda_3}{\varepsilon+P} \leq 1$,  $\Lambda_2 = - (\Lambda_1 + \Lambda_3)$.

\begin{figure}
    \centering
    \includegraphics[width=\linewidth]{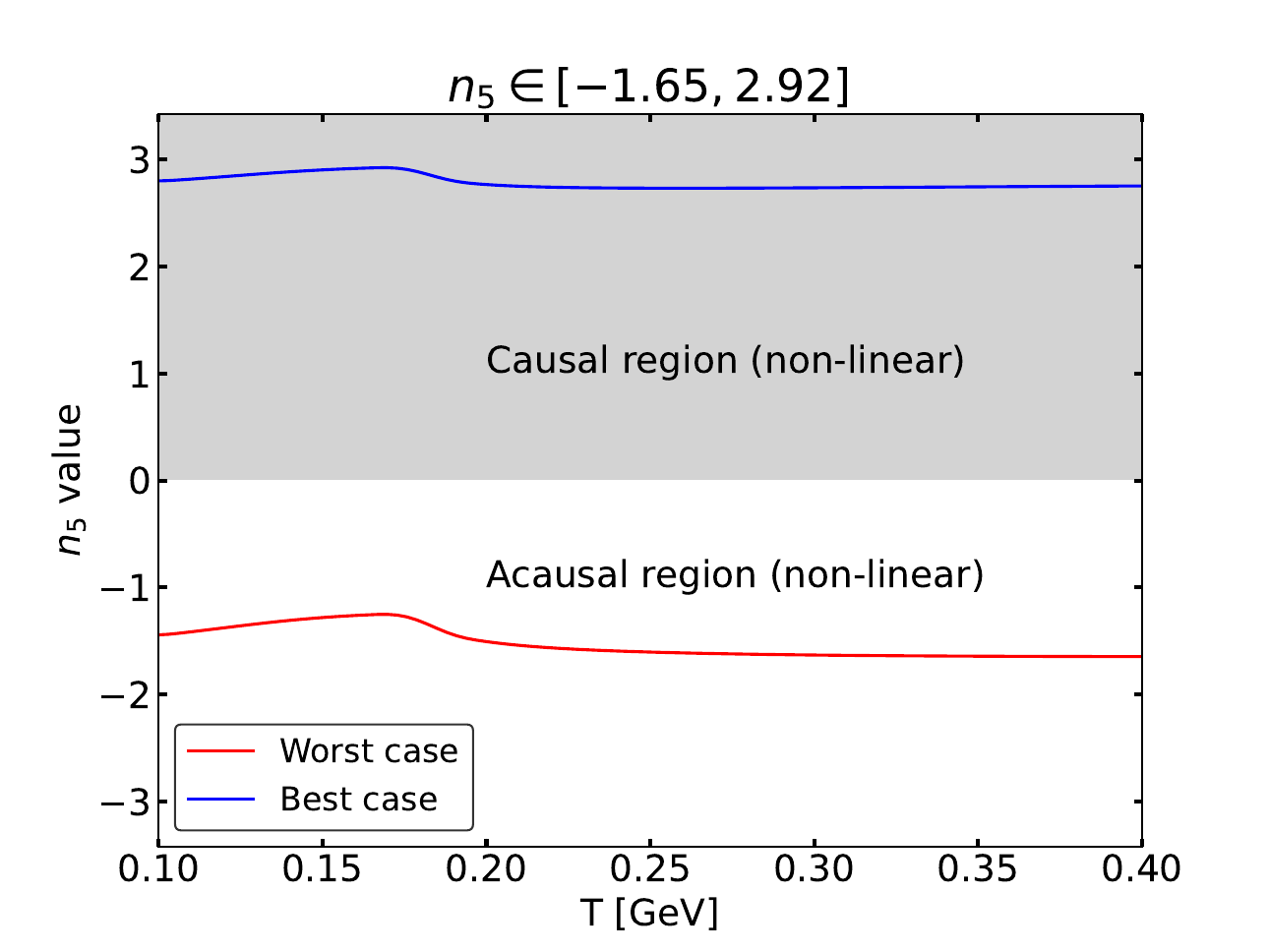}
    \caption{
    Representation of temperature-dependent causality condition $n_5$, \eqref{eq:n5}, along with ``best'' and ``worst'' cases defined by limiting inverse Reynold's numbers to be between zero and unity: $0 \leq \frac{\Pi}{\varepsilon+P} \leq 1$, $-1 \leq \frac{\Lambda_1}{\varepsilon+P} \leq 0$, $0 \leq \frac{\Lambda_3}{\varepsilon+P} \leq 1$,  $\Lambda_2 = - (\Lambda_1 + \Lambda_3)$.
    }
    \label{fig:n_5}
\end{figure}

\begin{figure}
    \centering
    \includegraphics[width=\linewidth]{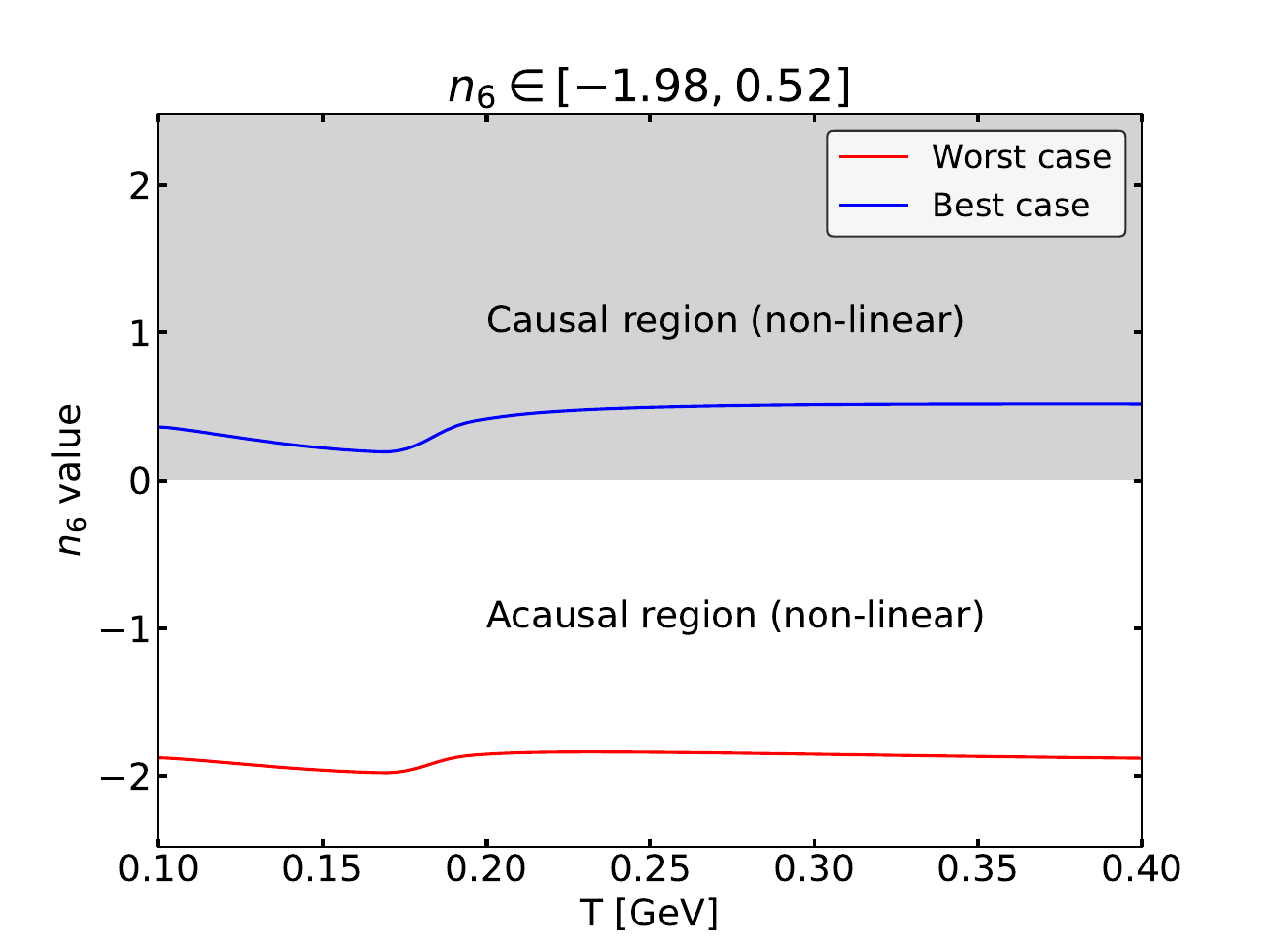}
    \caption{
    Representation of temperature-dependent causality condition $n_6$, \eqref{eq:n6}, along with ``best'' and ``worst'' cases defined by limiting inverse Reynold's numbers to be between zero and unity: $0 \leq \frac{\Pi}{\varepsilon+P} \leq 1$, $-1 \leq \frac{\Lambda_1}{\varepsilon+P} \leq 0$, $0 \leq \frac{\Lambda_3}{\varepsilon+P} \leq 1$,  $\Lambda_2 = - (\Lambda_1 + \Lambda_3)$.  This is typically the strictest and most important condition.}
    \label{fig:n_6}
\end{figure}

Using these limits, the resulting ranges for the first four conditions are
\begin{align}
    n_1 &\in [-0.46, 2.2]\\
    n_2 &\geq 0 \text{ always}\\
    n_3 &\in [-0.23, 1.1]\\
    n_4 &\in [-0.5, 1.875].
\end{align}
Since $n_5$ and $n_6$ depend on temperature through the speed of sound, we plot them versus temperature in Figures \ref{fig:n_5} and \ref{fig:n_6}, and we can see that these are the strictest conditions of the 6.   In particular, $n_6$ is the strongest condition, as has been noticed previously \cite{Chiu:2021muk,ExTrEMe:2023nhy}.
\subsubsection{Parameter Posterior}
\label{sub_sec:params_posterior}
In order to impose causality conditions on the Bayesian prior, we need to quantify the breaking of causality.   The simplest would be a simple binary --- if causality is ever violated at any point in spacetime, in any simulation at any centrality, then we could conclude that the corresponding model parameters are not allowed and the prior probability density at that point in parameter space should be zero.  However, while this ensures that the hydrodynamic equations are used in an acausal regime,
this may be overly strict.  If the fluid passes through a non-hydrodynamic regime in only a small part of its spacetime evolution, it is possible that the final simulation results are still close to their realistic values, and we may not want to completely discard the model at that point in parameter space.

Instead, we want to make a quantification of how much causality violation there is, so that we can put varying cutoffs to the amount of allowed violation, and see the corresponding effects on the final Bayesian posterior.

We first note that the largest number of acausal fluid cells typically occur at the earliest times, where gradients can become large. With this in mind, we analyze the system at the onset of hydrodynamic evolution (at the end of the free-streaming phase) and determine the spatial region that violates the nonlinear causal conditions, and the region that instead satisfies the inequalities.  We quantify the fraction of the system that is acausal as the fraction of the total initial energy that lies in an acausal region,
\begin{align}
    F_{\rm ac} &= \frac{E_{\rm acausal}}{E_{\rm total}}.
\end{align}
%

More specifically, we generate 20000 random samples of the original posterior from the JETSCAPE analysis of Pb-Pb data using the Grad model for viscous correction to the distribution function at particlization~\cite{JETSCAPE:2020mzn}.   For each of these points in parameter space, we perform one simulation at zero impact parameter of $\mathrm{T}_{\mathrm{R}} \mathrm{ENTo}$ plus free-streaming, and analyze the resulting hydrodynamic initial conditions to determine whether any of the necessary causality conditions are violated in each fluid cell and quantify the energy contained in these acausal regions as compared to the total energy.

\begin{figure}
    \centering
    \includegraphics[width=\linewidth]{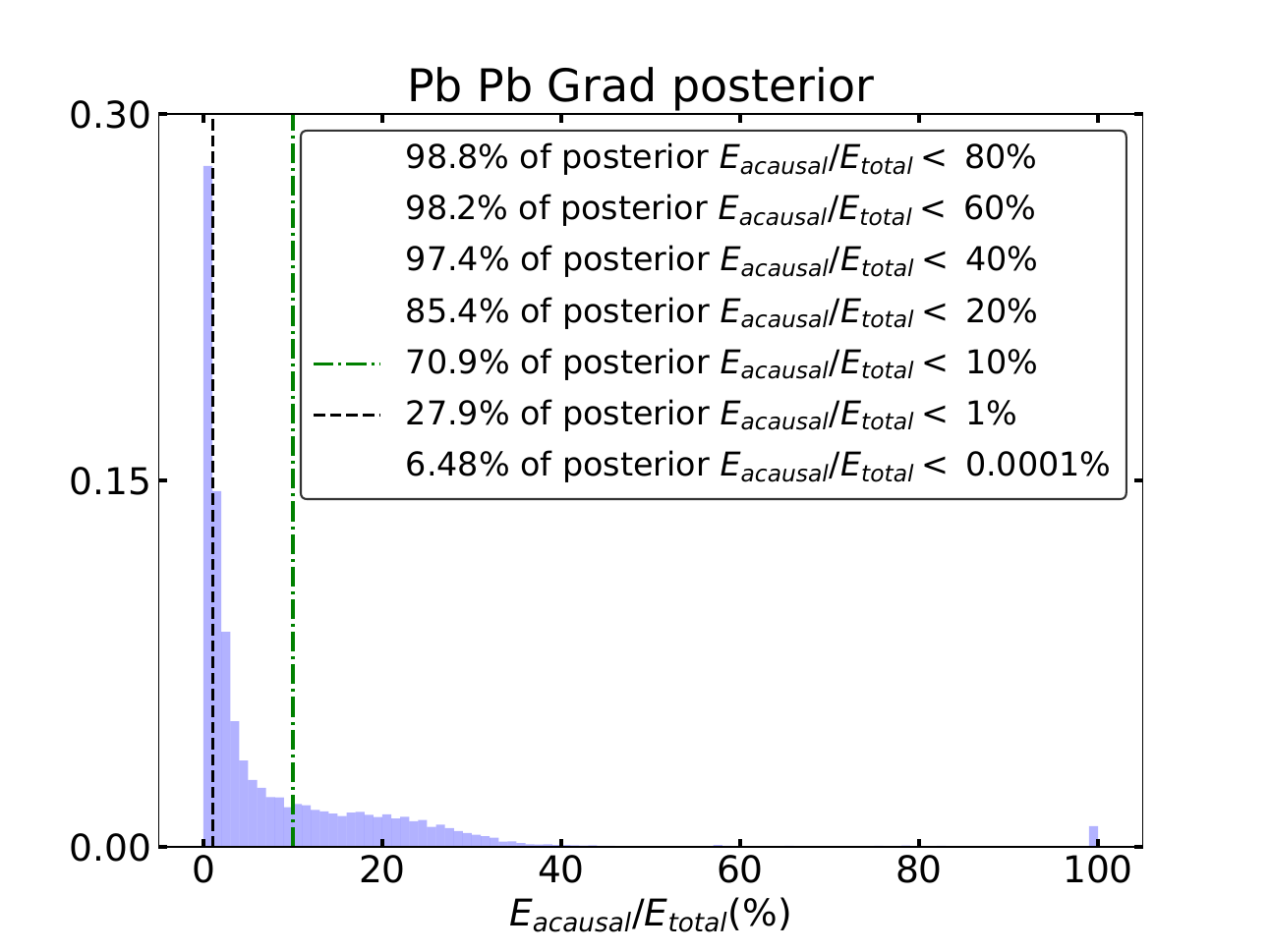}
    \caption{Histogram of the posterior distribution for causal energy fraction $F_{\rm ac}$ in a Bayesian analysis of Pb-Pb collisions at $\sqrt{s_{NN}} = 2.76$ TeV using Grad model for viscous corrections at particlization. Also shown are a number of percentile. This plot shows a large number of parameter space points in an acausal behavior, especially when $b_{\pi}$ is close to 2, and we have almost 100 \% of acausal energy where the system is completely acausal.}
    \label{fig:posterior_histogram}
\end{figure}

In Fig.~\ref{fig:posterior_histogram} we show the posterior distribution for acausal energy fraction.  We see that there is a high density at very small energy fraction,
indicating that a significant part of the model parameter space has a relatively small amount of acausality, in agreement with~\cite{ExTrEMe:2023nhy}.
Nevertheless, the distribution is fairly wide, with some parameter samples ($\sim$1\%) where close to 100\% of the system is initially in an acausal regime.  Other percentiles are also shown in Fig.~\ref{fig:posterior_histogram}, indicating that demanding the fraction of the system initially in an acausal regime necessarily be less than a certain amount (say 10\%), results in an exclusion of some amount of the previously-favored posterior parameter space (in this case $\sim$30\%).  
An important question, then, is if we place limits on the allowed acausal energy fraction in the model, is there an effect on the Bayesian posterior for physical parameters and observables?

Besides the relaxation times that appear in the linear case, the full nonlinear inequalities depend on local conditions --- in particular the bulk pressure and the shear viscous tensor.  The values of these dissipative quantities at the onset of hydrodynamic evolution depend on the pre-hydrodynamic model and the associated parameters ($N$, $p$, $w$, $d_{\rm min}$, $\sigma_k$, $\tau_R$, $\alpha$).  A restriction on acausality in the initial state therefore directly affects the prior (and therefore also the posterior) of these parameters.   Nevertheless, the allowed posterior of the other model parameters (such as viscosity) can still be significantly affected due to their correlation with initial-state parameters.  That is, a restriction of the initial-state model can change which hydrodynamic parameters are allowed or preferred.

We start by placing various cutoffs for the allowed acausal energy fraction $F_{\rm ac}$, and investigating the effect of removing the offending regions of parameter space from the Bayesian prior.  

\begin{figure}
    \centering
    \includegraphics[width=\linewidth]{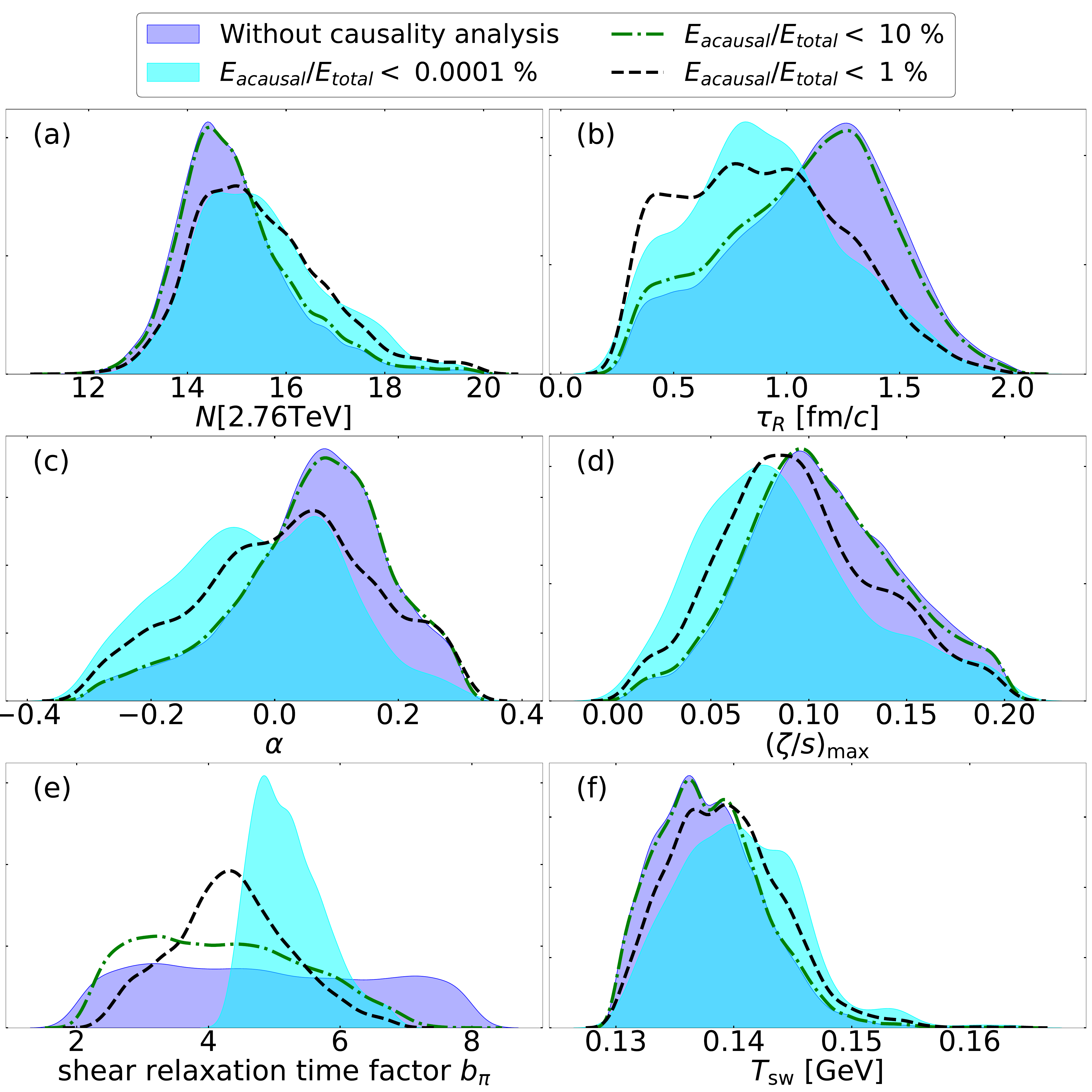}
    \caption{
    Marginalized posterior distributions for selected parameters with various allowed acausal energy fractions.  We note that a limit of $F_{\rm ac}<10\%$ removes 29\% of the total posterior parameter space, a cutoff of $F_{\rm ac}<1\%$ removes 72\%, and with a cutoff of $F_{\rm ac}<10^{-4}$\% only 6.5\% of the original posterior remains (see Fig.~\ref{fig:posterior_histogram}). The panels represent the following parameters: a) $N$ normalization in the $\mathrm{T}_{\mathrm{R}} \mathrm{ENTo}$ model, b) and c) parameters for the free-streaming time given in the equation \ref{eq:taufs}, d) peak of the parametrization given for the bulk viscosity \ref{eq:bulk_parametrization}, e) shear relaxation time factor given in equation \ref{eq:tau_pi} and f) switching temperature.
    }
    \label{fig:params_posterior}
\end{figure}

In Fig.~\ref{fig:params_posterior} we show the marginalized posterior for a selection of parameters that are most affected by restrictions on causality violation.  The 1D marginalized posterior for every parameter and the 2D posterior for every possible pair of parameters can be found in the Appendix.
%
While the linear causality condition is only violated for small values of the relaxation time factor $b_\pi$, our results show that large values can also lead to acausality. The odd-numbered nonlinear causality conditions include a positive term proportional to $1/C_\eta = 1/b_\pi$, while the even-numbered ones feature a negative term. Therefore, both very large and very small values of $b_\pi$ can be problematic, particularly in conditions $n_5$ and $n_6$. When the viscous tensor is large, $n_5$ can become negative for large $b_\pi$, but becomes positive as $b_\pi$ decreases and $n_6$ is a nonlinear extension case of the linear causality condition. Thus, the relaxation time must remain within a balanced range to maintain relativistic causality in realistic simulations, as noted in \cite{ExTrEMe:2023nhy}.

Next, we investigate the marginalized posteriors of initial-state parameters.  In particular, we find a significant effect on the free-streaming time parameters $\tau_R$ and $\alpha$, also shown in Fig.~\ref{fig:params_posterior}.
In general, demanding strong constraints on causality in the initial hydrodynamic evolution preferentially selects for smaller free-streaming time, and an earlier onset of hydrodynamics (though not too early).   Noting that earlier times generally correspond to stronger longitudinal expansion and therefore might be expected to have larger deviations, this may seem unexpected.  However, later start times also correspond to lower temperatures, closer to the transition temperature, where the strictest condition $n_6$ is itself the most stringent. Combined with the large bulk pressure and shears present in the (conformal) pre-hydrodynamic model \cite{Borghini:2024kll}, causality demands that the hydrodynamic phase not start too late. Similarly, we find that the normalization factor $N$ shouldn't be too small, likely for the same reason.

In Fig.~\ref{fig:params_posterior} we also show a final-state parameter, the switching temperature $T_{\rm sw}$ where hydrodynamic degrees of freedom are switched for particle degrees of freedom before being sent through the hadronic cascade.   We can see a slight preference for larger switching temperatures if strong constraints on causality are imposed.

\begin{figure}
    \centering
    \includegraphics[width=\linewidth]{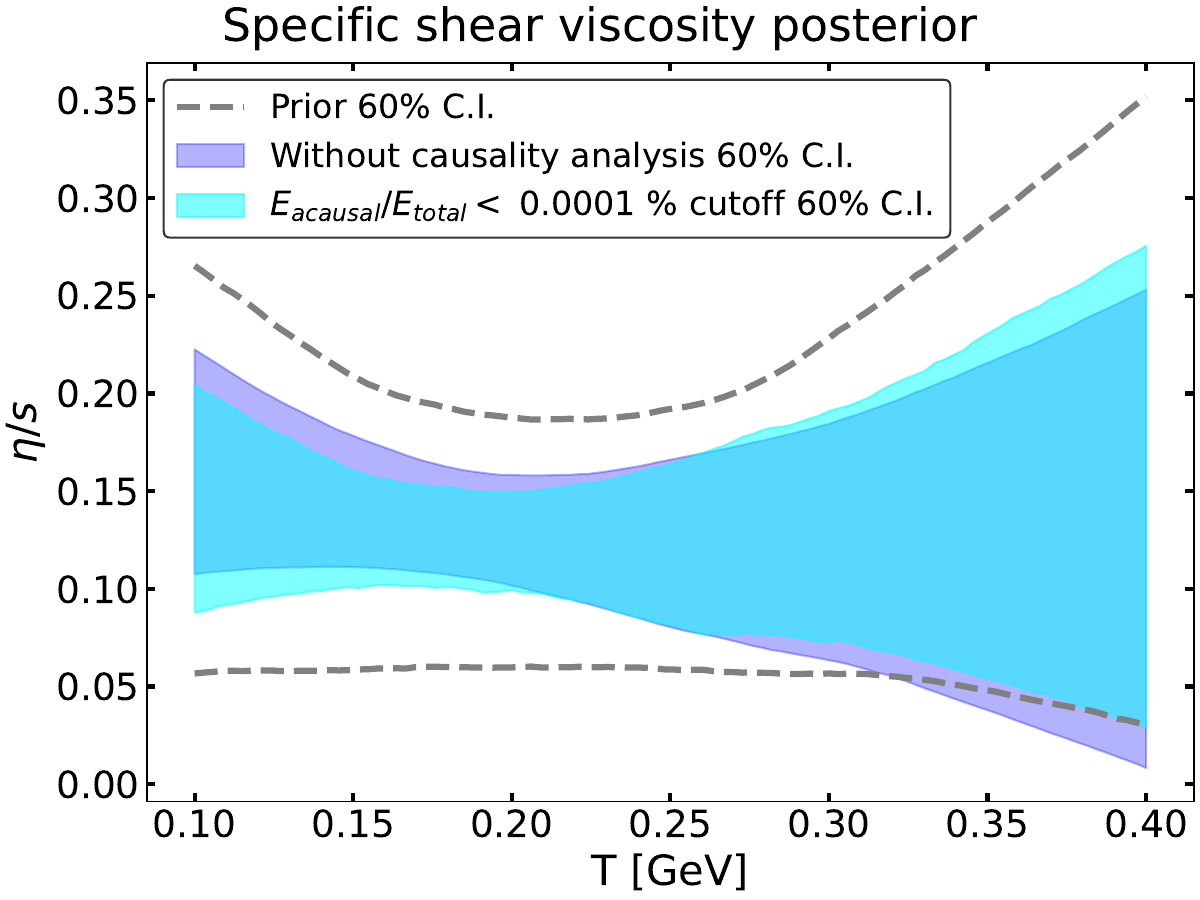}
    \includegraphics[width=\linewidth]{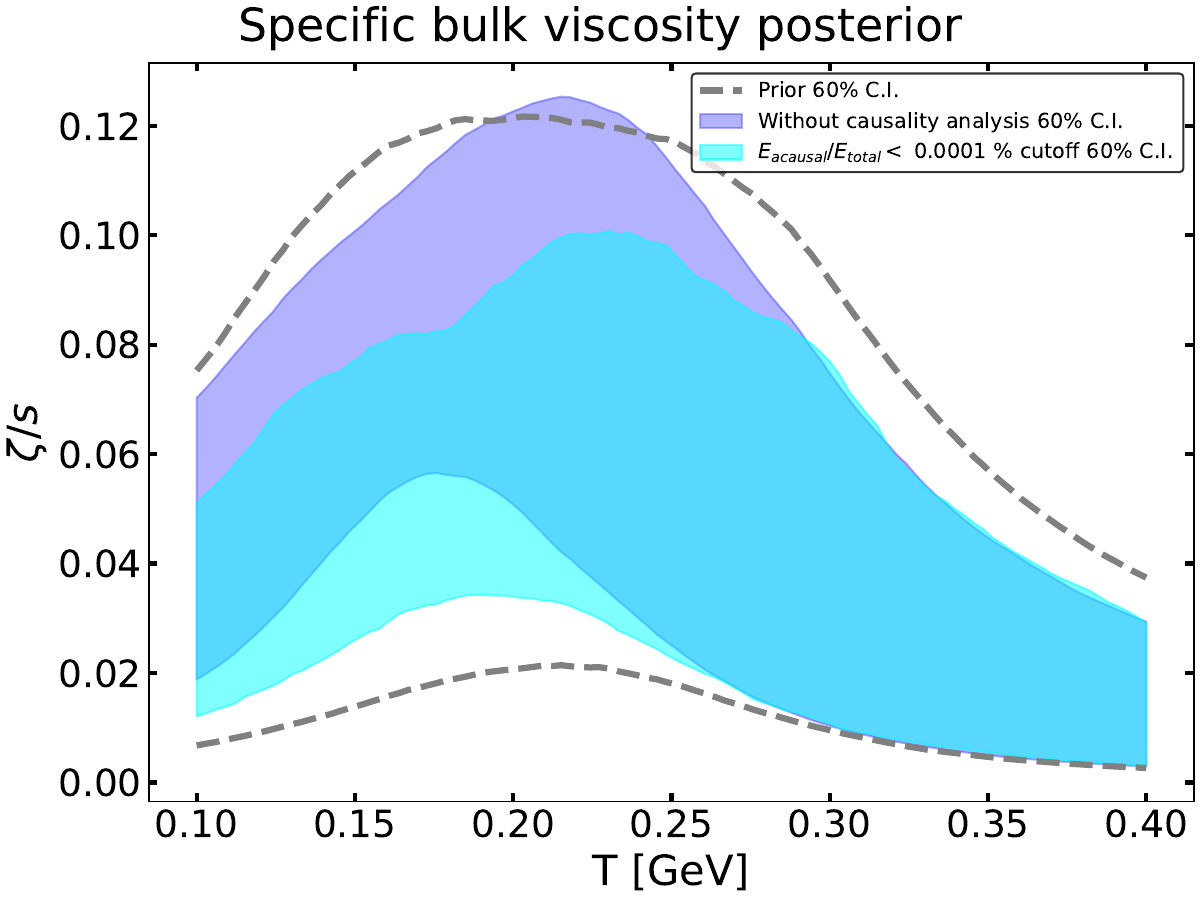}
    \caption{Specific shear and bulk viscosities as a function of temperature.  Here we show the 60\% credible interval for the full prior, the full posterior with no causality restriction, and when the initial acausal energy fraction is not allowed to exceed $10^{-4}$. We can see in particular that bulk viscosity is significantly biased by acausal regimes.}
    \label{fig:shear_and_bulk_posterior}
\end{figure}

Fig.~\ref{fig:shear_and_bulk_posterior} shows the difference in the posterior for bulk and shear viscosity imposing the causality conditions, allowing a maximum of $10^{-4}\%$ of acausal energy fraction. It is possible to notice the specific bulk viscosity is more affected by this imposition. 
\subsubsection{Observables}
\label{sub_sec:observables}
It is interesting to ask how the restricted parameter posterior affects the observable posterior.  That is, to what extent does the removal of certain parts of parameter space inhibit the ability of the model to describe experimental data?  

\begin{figure}
    \centering
    \includegraphics[width=\linewidth]{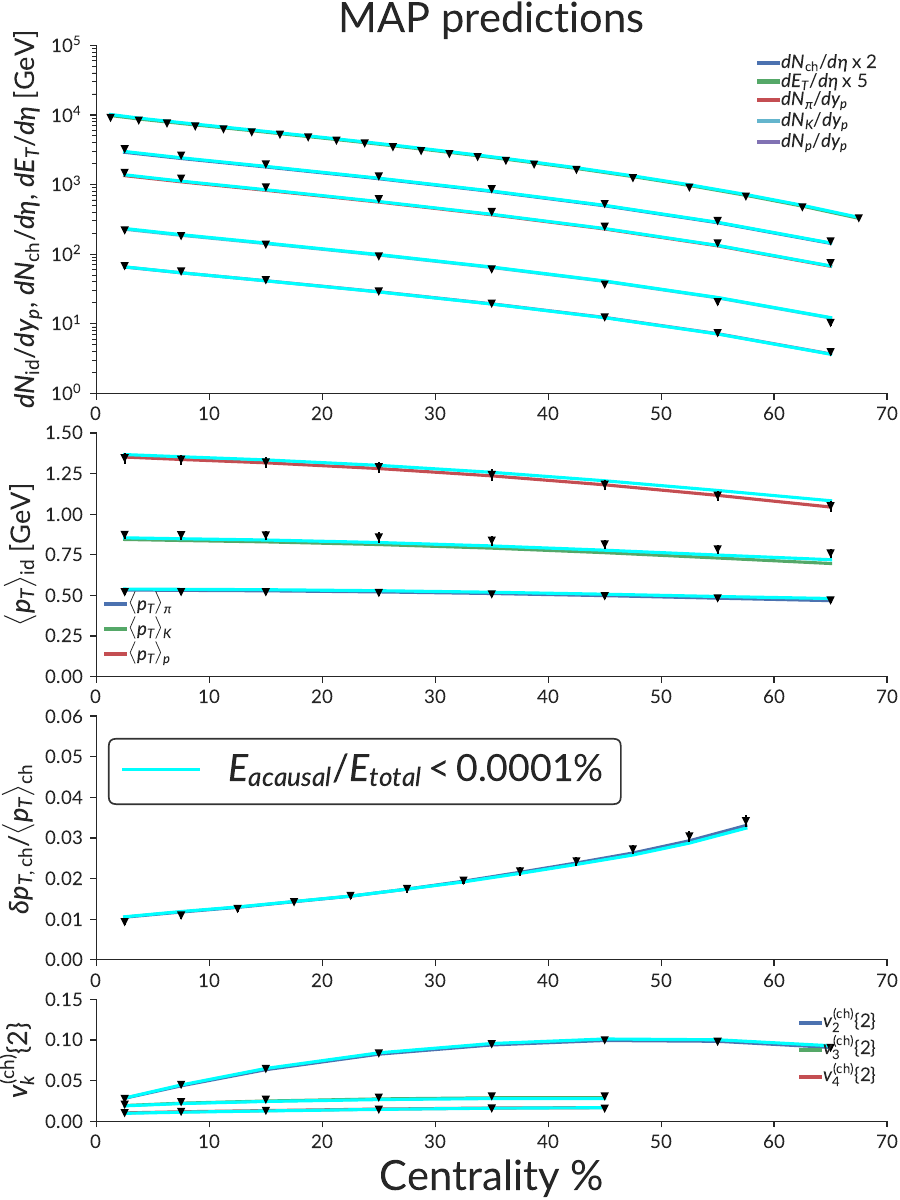}
    \caption{
    Posterior observable predictions for the Maximum A Posteriori (MAP) parameter values.  Shown are the original MAP predictions from JETSCAPE with no restrictions on causality (solid colored lines)~\cite{JETSCAPE:2020mzn} and the MAP after a restriction of $F_{\rm ac} < 10^{-4}$ (dashed aqua lines).
    }
    \label{fig:observables_fit}
\end{figure}

In Fig.~\ref{fig:observables_fit}, we show the observable prediction for the Maximum a Posteriori (MAP) parameter set.  These are the parameter values that give the best possible simultaneous fit to experimental data as defined by the maximum of the posterior probability distribution.   We show the result for the original, unconstrained analysis (as estimated by 20000 posterior samples), and with an acausality fraction cutoff of $F_{\rm ac} \leq 10^{-4} \%$ (as estimated by the remaining $\sim$1300 samples).  

The fit to experimental data necessarily worsens when constraints are imposed on the model --- specifically, the posterior probability is reduced to approximately 28\% of the pre-cut value -- that is, in the original analysis such a parameter value would be less probable by more than a factor 3, compared to the probability of the best value.   Nevertheless, the fit is not degraded so much so as to prevent the general ability of the model to describe data, with no significant discrepancies apparent by eye, in agreement with previous works which showed no difference in final observables for parameters that improve causality \cite{ExTrEMe:2023nhy, Chiu:2021muk}.

\section{Conclusions}
\label{sec:conclusions}
In this work, we analyze recent Bayesian analyses of heavy-ion collision data using new information about the validity of the underlying model.  Specifically, we use the violation of relativistic causality at the onset of hydrodynamic evolution to penalize or remove regions of parameter space that lead to such violation.   We confirm previous findings that typically-favored parameter values lead to the hydrodynamic equations being used outside their regime of validity in at least some fraction of the system evolution. We find that restricting the acausal parameter space significantly changes posterior parameter distributions, and therefore the inferred value of physical properties such as bulk viscosity of the quark-gluon plasma, which now disfavors the large values that were previously favored.

These results illustrate the importance of careful study of the regime of validity of simulations models, and especially the development of better models for the far-from-equilibrium QCD dynamics that occurs in the early stages of a relativistic nuclear colliison.

\section*{Acknowledgments}
We thank Dekrayat Almaalol and Jacquelyn Noronha-Hostler for the useful discussions, and we thank the JETSCAPE Collaboration for sharing simulation data.  T.S.D acknowledges financial support from CNPq grant
number 131740/2021-0 and INCT-FNA research project number 464898/2014-5. 
M.L.~was supported by FAPESP projects 2017/05685-2 and 2018/24720-6,  by project INCT-FNA Proc.~No.~464898/2014-5, and by CAPES - Finance Code 001. R.K. acknowledges support by the Deutsche Forschungsgemeinschaft (DFG, German Research Foundation) through the CRC-TR 211, project number 315477589 - TRR 211. J.N.  is partly supported by the U.S. Department of Energy, Office of Science, Office for Nuclear Physics under Award No. DE-SC0023861. J.-F.~P.~is supported in part by the U.S. Department of Energy, Office of Science under Award Number DE-SC-0024347. T. N.dS. is supported by CNPq through the INCT-FNA grant 312932/2018-9 and the Universal Grant 409029/2021-1.

\section*{Appendix: Full posterior of model parameters}
\begin{figure*}
    \centering
    \begin{minipage}[!]{1\linewidth}
    \includegraphics[width=\linewidth]{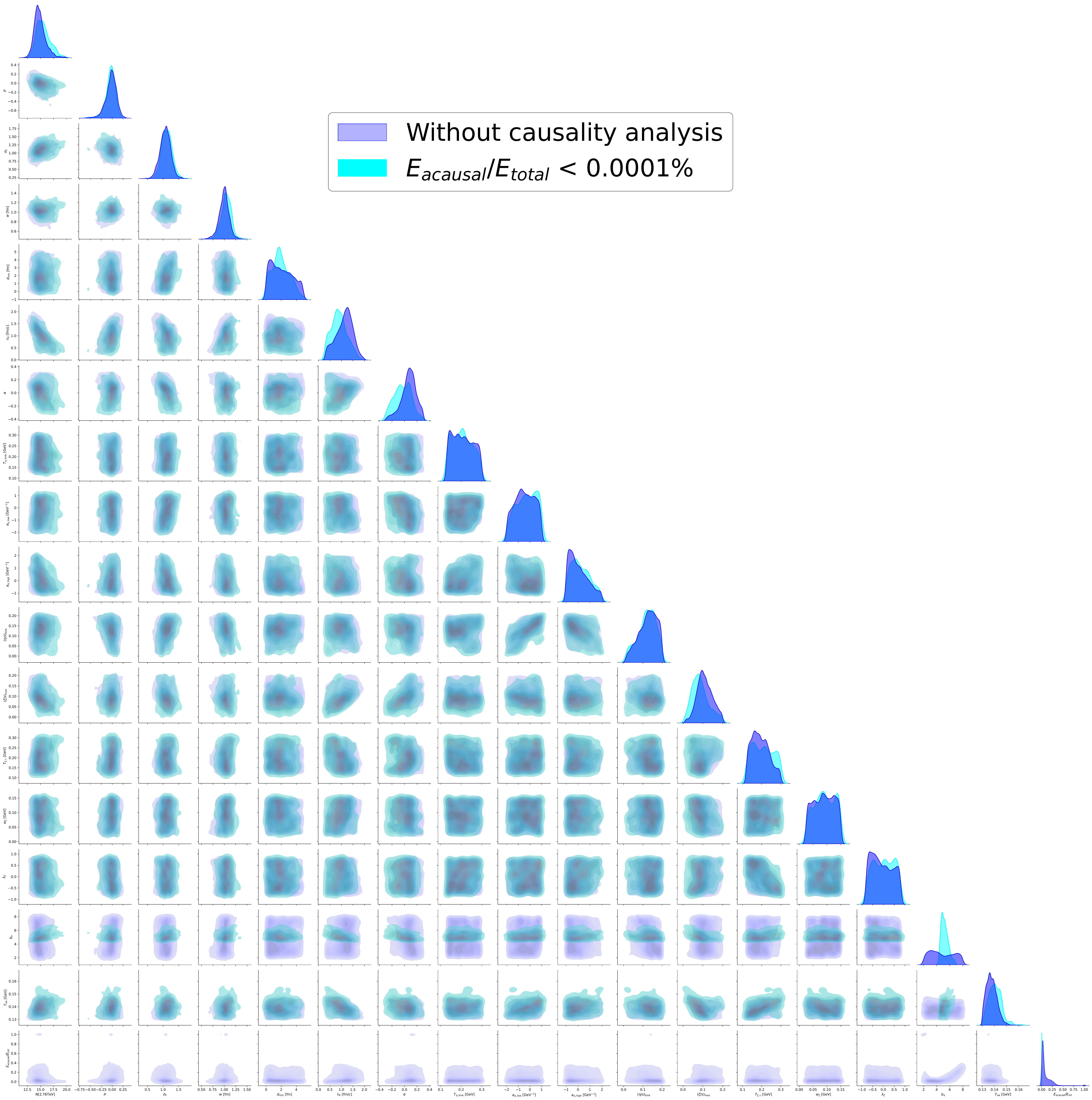}
    \end{minipage}
    \caption{Marginal distribution. Off-diagonal plots: pairwise (joint) probability distribution showing the correlations between the model parameters Diagonal plots: single probability distribution}
    \label{fig:marginalizations}
\end{figure*}

For completeness, we show in Fig.\ref{fig:marginalizations} the posterior distribution of all model parameters single and joint-parameter marginal distributions for the Grad viscous correction model (blue) and using the restriction of $F_{\rm ac} < 10^{-4}$ (aqua).    

\bibliography{ref.bib}
\end{document}